\documentstyle[12pt]{article}

\begin{document}

\title{The Classical and Quantum Mechanics of Lazy Baker Maps}
\date{}
\author{A. Lakshminarayan and  N.L. Balazs \\ {\sl Department of Physics,
State University of New York at Stony Brook}\\
{\sl Stony Brook, N.Y. 11794}}

\maketitle

\newcommand{\newc}{\newcommand}
\newc{\beq}{\begin{equation}}
\newc{\eeq}{\end{equation}}
\newc{\beqa}{\begin{eqnarray}}
\newc{\eeqa}{\end{eqnarray}}
\newc{\longra}{\longrightarrow}
\newc{\longla}{\longleftarrow}
\newc{\ul}{\underline}\newc{\rarrow}{\rightarrow}
\newc{\br}{\langle}
\newc{\kt}{\rangle}
\newc{\hs}{\hspace}
\begin{abstract}

We introduce and study the classical and quantum mechanics of certain
non hyperbolic maps on the unit square. These maps are modifications
of the usual baker's map and their behaviour ranges from chaotic
motion on the whole measure to chaos on a set of measure zero. Thus we
have called these maps ``lazy baker maps.'' The aim of introducing
these maps is to provide the simplest models of systems with a mixed
phase space, in which there are both regular and chaotic motions. We
find that despite the obviously contrived nature of these maps they
provide a good model for the study of the quantum mechanics of such
systems.  We notice the effect of a classically chaotic fractal set of
measure zero on the corresponding quantum maps, which leads to a
transition in the spectral statistics. Some periodic orbits belonging
to this fractal set are seen to scar several eigenfunctions.

\end {abstract}
\newpage

\section{Introduction}
           \hs{.2in} We introduce a class of maps of the unit square
that can be given the topology of a torus. These maps are
discontinuous, non-hyperbolic, area preserving, piecewise linear and
include the geometric actions of stretching, folding and rotation.
The disparate dynamics due to these actions result in a wide range of
dynamic behaviour, ranging from full fledged chaos to exactly periodic
motion on the whole measure. The purpose of introducing these maps is
that they can be easily quantized, and thus provide some of the {\em
simplest quantum models of systems that are not completely
hyperbolic}.  The classical maps in themselves have also proved to be
very interesting, even though, as yet, they have no immediate physical
applicability.

Previously several maps of the torus onto itself have been studied
with a view to adopt them to quantum mechanics. The class of maps
called cat maps [1] have proven to have a non-generic quantum
mechanics, while the other much studied model of the standard map [2]
has the drawback that information on periodic orbits and the other
classical features are hard to come by. The classical baker's map
which has also been quantized is completely hyperbolic [7]. The maps
studied here are the simplest manifestations of chaos and order
existing in the same system. This class of problems has been
recognised as important and describe many physical systems where not
all KAM torii have been destroyed.

	The quantization of the completely hyperbolic baker's map has
proven to be a very useful model of quantum chaology, and the
following maps should be considered as modifications of the baker's
map.  They range from the action of bakers who thoroughly mix up the
dough (phase fluid ) to completely lazy bakers who do not mix the
dough at all.

\section{ $SRS$ Maps}
     \hs{.2in}	We denote by $SRS$ the class of maps defined
geometrically by dividing the fundamental unit square into 3 vertical
rectangles and stretching the first and third and rotating the second
about the center of the square into horizontal rectangles as shown in
fig.(1). We denote by $S_{1}(a)$ the symmetric maps where the first
and third vertical rectangles are of equal measure ($a=1-b$).  Here
$a$ and $b$ are the locations of the first and second cut on the
horizontal axis; hence the area of the first and third vertical
rectangles are $a$ $(0<a<1/2)$.  $S_{1}(a)$ are chaotic only on a set
of measure zero, and are critical maps in a sense we discuss below.
The breaking of symmetry seems to produce ergodic behaviour on a set
of finite measure. In this note we do not study this facet of the
problem, except through the numerical results of the
corresponding quantum models.

\subsection{ Classical Mechanics of $SRS$ Maps}

\hs{.2in} The $S_{1}(a)$ map is not even ergodic, let alone mixing. We employ
simple symbol sequences to extract the dynamics. The map is given by

\[ \left. \begin{array}{ccc}
        q ^{ \prime}  & =& q/a \\
        p ^{ \prime}  & = & p a             \end{array}
            \right\} \mbox{     if} \; \; 0 \leq q \leq a
\]
\beq  \left. \begin{array}{ccc}
         q ^{\prime}  & =& 1-p \\
         p ^{\prime} & = & q
            \end{array}
            \right\} \mbox{      if} \; \;  a < q < 1-a
\eeq
\[ \left. \begin{array}{ccc}
         q ^{\prime}  & =& (q-1+a)/a \\
         p ^{\prime}  & = & pa +1-a
            \end{array}
            \right\} \mbox{    if}  \; \; 1-a \leq q \leq 1  \]

\begin{sloppypar}
We now analyse the map $S_{1}(1/3)$. Later we show that all the
symmetric maps $S_{1}(a)$ are topologically conjugate to each other.
This means that their classical dynamics are essentially identical due
to the existence of a one to one transformation between the two maps
which commutes with time [5]. We use the ternary representation for $q$
and $p$; thus we represent a state in the square by a bi-infinite
sequence with the left side specifying the position and the right side
specifying the momentum, both being in their ternary basis. The
symbols 0,1 and 2 now represent the first second and third equal
vertical or horizontal partition of the unit square. The map's
dynamics translates into the following rules on the bi-infinite
sequence.  If the most significant bit in the position is $0$ or $2$
the dynamics is a left shift, but if the most significant bit is $1$,
position and momentum get interchanged and the new position bits ( or
old momentum bits) have their $0$ and $2$ exchanged (two's complement
). For example \\
\[ \cdots12210.02101022\cdots \rarrow \cdots122100.2101022\cdots\]

\[\rarrow \cdots1221002.101022\cdots\rarrow \cdots220101.0221001.\]
\end{sloppypar}

The dynamics is one of pure left shift only for those momenta and
positions between $0$ and $1$ that have only $0$ and $2$ in their
ternary basis representation, this, of course, being the standard
middle third Cantor set ($C_{1/3}$) [3].  Therefore the union of $
C_{1/3} \times [0,1] \;\;
\mbox{and} \; \; [0,1]\times C_{1/3}$, is chaotic. We can easily show that
the {\em complement of the union of these two sets consists entirely of
periodic orbits, and since the Lebesgue measure of the middle third
Cantor set is zero this map and its relatives are extremely regular}.
The chaotic set in $S_{1}(1/3)$ is a fractal, whose Hausdorff
dimension is $1+Log2/Log3$ [3]. We refer to this set below as the
``fractal set'', the ``hyperbolic set'' or the ``fractal hyperbolic
set'' of the map.

Any state with at least one $1$ bit in both position and momentum must
traverse the rotating third of phase space. This allows us to follow
the fate of states like
\beq \mbox{anything}\:1a_{m}a_{m-1}\cdots a_{1}.1\: \mbox{anything}
\eeq
where the anythings represent any infinite string comprising of $0$,
$1$, and $2$.  Let
\[\gamma_{m} \equiv a_{1}a_{2}\ldots a_{m}, \; \; a_{i}\in \{0,1\},
m=0,1,2,\ldots.\]
It is easy to see that the above state represents a rectangular island
(the $\gamma$ island) of area $3^{-m-2}$ and that this island as a
whole is periodic(!) with period $4(m+1)$. (See
Appendix A). There are isolated points within these islands which have
period $(m+1)$ (if m is even) or $2(m+1)$ .They represent elliptic
periodic orbits with rational rotation numbers and inverse parabolic
orbits respectively; actually they are just either rotation by
$\pi/2$ or $\pi$. The rest of the periodic orbits of the $\gamma$
island are direct parabolic. Thus periodic orbits occur in continous
families reminiscent of integrable systems.  These dynamical systems
are non-generic and have not been previously discussed. The main
feature of the transformation ( say $T$) is that for every integer $n$
there exists regions of finite measure $\sigma$ such that
$T^{4n}(\sigma)=1$, and $n$ is the least such integer.

To counter the claim that this trivialises the system we must note
that the map has {\em periodic orbits of all periods and a chaotic set
of measure zero}. This chaotic set effectively becomes measure zero
only at {\em infinite times} in the sense that the fractal set develops
with increasing time.  We know from previous studies [7,8,9] that finite time
orbits, especially the short periodic orbits, seem to strongly influence the
spectra and eigenfunctions for a finite, albeit small Planck's
constant. Which means that in {\em quantum mechanics} we may expect
this measure zero set to play a role.  That this is in fact the case is
borne out by the later discussion of the quantum $SRS$ map (Sec.3).

The $\gamma$ island is part of an island chain amongst which the
dynamics is closed and periodic. Given $m$ and hence the period, there
are\\
\beq \left. \begin{array}{ll}
           2^{m-1} & m \;\mbox{ odd}\\ 2^{m/2-1} (1+2^{m/2}) & m \;
\mbox{ even} \end{array} 	\right.
\eeq
such distinct island chains. (See Appendix A). Further each island chain
consists of $2(m+1)$ ($m$ odd or $m$ even and $\gamma \neq
\bar{\gamma}^{\dagger}$) or $(m+1) $ ($m$ even and $\gamma =
\bar{\gamma}^{\dagger})$ number of islands.  We use an overbar to
denote the complement operation and the dagger to denote
``conjugation'' or reflection, which we define below. Here
$\bar{\gamma}^{\dagger}$ is the $2$'s complement of the string
$\gamma$ compounded with a reflection. If the string length is $m$
then a reflection is defined as $a_{l} \, \rarrow \,a_{m-l+1}$.
Therefore the total number of period $4(m+1)$ islands is $2^{m}(m+1)$
irrespectively of, $ m $ being even or odd.

 Since each island is of area $3^{-m-2}$ and \[ \sum_{m=0}^{\infty}
2^{m}(m+1) 3^{-m-2} = 1\] the Lebesgue measure of direct parabolic
periodic orbits is unity and the periods are all multiples of 4.

\subsection{Reducing the area of rotation}

\hs{.2in} We now want to study the effect of reducing the area
 of rotation while preserving the $R$-symmetry of the map. $R$
symmetry is defined as the invariance of the equations of motion under
the following transformation which is a reflection about the center of
the square:
\beq
      \begin{array}{ccc} q & \longra & 1-q \\ 	p & \longra & 1-p .
\end{array}
\eeq
We can do this by simply increasing $a$, or by introducing more
vertical partitions.  Increasing $a$ gives us a set of maps that are
topologically conjugate to each other. This implies that they share
identical orbit structures, such as periodic orbits and chaotic sets,
even though the measures of these sets might be different.  The
required homeomorphism ( a strictly continous one to one function) $f$
should commute with the dynamics of $S_{1}(a)$ and $S_{1}(1/3)$. Thus if
$T$ denotes the transformation $S_{1}(1/3)$ and $T^{\prime}$ the
transformation $S_{1}(a)$ then the following must be true:
\beq
               f \, \circ \, T = T^{ \prime} \, \circ \, f .
\eeq

Such a function is one that maps the middle third Cantor set into the
middle ``(1-2a)-th'' Cantor set (defined similar to the middle third set)
with the intermediate points being interpolated by straight lines
joining the immediate neighbouring points belonging to the Cantor
sets. This homeomorphism applied to the position and momentum makes
the dynamics of $S_{1}(a)$ commute with that of $S_{1}(1/3)$, and
hence these maps are topologically identical. $f$ is a Cantor function
and is a dynamically generated homeomorphism of the kind previously
discussed in ref.[4]. Such a function is one to one and monotonously
increasing with a countably infinite number of points where the
derivative is not defined. The scaling properties that makes $f$ the
required homeomorphism are
\beq \left. \begin{array}{lccc} f(q/a) & =& 3\, f(q)&
\;\; \;0 \leq q \leq a \\ f((q-1+a)/a)& = & 3\, f(q) -2 & \; \; \;
1-a\leq q \leq 1. 	\end{array} \right.
\eeq

The periodic orbit structure of $S_{1}(a)$ is therefore identical to
that of $S_{1}(1/3)$, and the chaotic set is of measure zero.  The
{\em quantum spectra, however, are different, and this is due to the changing
stability of short periodic orbits belonging to the hyperbolic fractal set that
for some time ``pretends'' to be in a sea of chaos}.

Increasing the number of partitions while keeping them all equal
produces classically simple dynamical systems.  So we consider
$S_{k}(1/l), k=1,2,3,...$ ($2k+1=l$ say). We divide the fundamental square
into $l$ vertical rectangles and stretch and fold all rectangles except the
central one which we choose to rotate. The extensions from $S_{1}(a)$ are
immediate. The chaotic set in the map is
\beq
C_{1/l} \times[0,1] \bigcup [0,1] \times C_{1/l}
\eeq
 where  $ C_{1/l}$  is the standard one {\sl l} th Cantor set. We illustrate
it's construction with $ l=5 $ as follows. Take the unit interval
and divide it into $5$ equal pieces and remove the central piece. The
next step is to divide each of the remaining $4$ pieces into $5$ more
equal pieces and removing the central piece from each. This process
repeated infinite number of times leaves behind a set of Lebesgue
measure zero, the  set $C_{1/5}$. The dynamics on the
complement of \[ C_{1/l} \times[0,1]
\bigcup [0,1] \times C_{1/l}  \]
again splits into periodic islands of period $4(m+1)$. There are
a total of $(2k)^{m}(m+1)$ islands of area $(2k+1)^{-m-2}$. The area of period
$4(m+1)$ islands therefore being $(2k)^{m}(m+1) (2k+1)^{-m-2}$ and we
see that shorter periodic orbits have relatively lower measures in
the higher $k$ maps.

\subsection{Breaking of $R$-Symmetry}
            \hs{.2in} The peculiar regularity of the maps $S_{1}(a)$
with a thin set of chaotic orbits is supported by the $R$-symmetry of
these models. The breaking of this symmetry ( when $a \ne 1-b$) leads
to the elliptic orbits and inverse parabolic orbits changing into
reflecting hyperbolic ones; it also seems to produce ergodic behaviour
on a set of non-zero measure. One can establish the ranges of the
parameters $a$ and $ b$ such that this change would happen  at least to the
large islands. These were originally direct parabolic of period 8, with
an inverse parabolic period 4 orbit at the center.  These maps are
themselves interesting and are most probably not structurally stable,
but their further discussion here is not warranted. It can be readily
seen that toroidal boundary conditions are compatible with the
breaking of this symmetry.

We show some results of computer iterations in fig.(2) where the
transformation of the parabolic island into a hyperbolic region is
evident. The extreme case in which the measure of the first partition
is one half the phase space and the second partition is a rotation
about the centre and the third partition has measure zero ($a=1/2,
b=1)$ is completely chaotic (We call this map $SR$). Such a map however
cannot be given the topology of a torus due to the disparate time
behaviour of the boundaries; in considering the quantum mechanics of
such maps we will therefore allow a thin third vertical partition.  We
now give some details on the map $SR$ that is chaotic on the whole
measure.

  	 The map $SR$ is given by
\beq \left. \begin{array}{ccc}
        q ^{ \prime}  & =& q/2 \\
        p ^{ \prime}  & = & 2\,p
            \end{array}
            \right\} \;\;\mbox{if} \; \; 0 \leq q \leq 1/2
\eeq
\beq \left. \begin{array}{ccc}
         q^{ \prime}  & =& 1-p \\
         p^{ \prime} & = & q
            \end{array}
            \right\} \;\;\mbox{if} \; \;  1/2 < q \leq 1.
\eeq
 It is ergodic, has a dense set of periodic orbits, and is sensitively
dependent on the initial conditions. It therefore satisfies the
requirements in the definition of chaos [5]. The positive Lyapunov
exponent is $ \log 2 \,/2 $ almost everywhere, and its topological
entropy is 1.46. The number of periodic orbits of period $T$ is
\beq p(T) = c_{1}a_{1}^{T}+c_{2}a_{2}^{T}+c_{3}a_{3}^{T}, \eeq

where $a_{1},\; a_{2},\; a_{3}$ are roots of the polynomial
equation
\beq a^{3} \; -a^{2} \; -1 \;=0 \eeq
and the constants $c_{1},\; c_{2},\; c_{3}$ are determined from
$p(3)=4,\; p(4)=5, \; p(5)= 6 $. This counts the fixed point at
$(0,0)$ but not the period $2$ orbit at $(1,1/2)$. All the periodic
orbits are hyperbolic and are either reflecting or ordinary. We
present some proofs in the Appendix B.

It is debatable if any self respecting baker, however lazy, will adopt
any of the $SRS$ maps discussed above. Our analysis of the $S_{1}(a)$
map strongly advises the baker against this map as mixing only a
measure zero subset of the dough can hardly be expected to produce
good pastries. However the intention of introducing these maps is an
attempt to quantize the simplest possible mixed systems.  The
semi-classics of these maps can then provide a testing ground for more
generic systems. The chaotic set of measure zero in $S_{1}(a)$ maps
may prevent us from proceeding to the strict semi-classical domain,
but may provide us a simple mixed system at finite, albeit small values of
the Planck's constant.

\section{Quantum $SRS$ Maps}

 \hs{.2in} We therefore turn to the quantum mechanics of these measure
preserving maps, which are unitary operators on a finite dimensional
vector space, having all the classical symmetries and possessing the
correct classical limit.

\subsection{The Propagators}

The  quantum mechanical one step propagator for $S_{1}(a)$ in the
position representation is given by:

\beq  B_{1} = G_{N}^{-1} \left( \begin{array}{ccc}
                         G_{Na} & 0 & 0\\
                           0 & I_{N(1-2a)} & 0 \\
                         0 &   0  & G_{Na}
                         \end{array}
                           \right) \eeq
where
\beq (G_{N})_{mn} = \br m|n \kt= \frac{1}{\sqrt{N}} \exp[-2 \pi i
(m+1/2)(n+1/2)] \,\,\,m,n=0,\ldots,N-1\eeq
 is the discrete Fourier transform matrix on $N$ sites and $I_{N(1-2a)}$ is
the $N(1-2a) \times N(1-2a) $ identity matrix. $N \, a$ is required to
be an integer and so the classical cut is allowed only at rational
points. The notation of inner product indicates that it is the
transformation matrix between discrete position states $|n\kt$ and
discrete momentum states $|m \kt$. These eigenstates are repeated
anti-periodically, that is

\beq |n \kt = - |n+N \kt. \eeq
 This results in the discrete Fourier matrix on $N$ sites, the entries being
shifted by $1/2$ in the exponent factors to allow the classical
$R$-symmetry to be preserved on quantization [8].

For details on such a quantization see the original quantization of
the baker's map in [7] and in [8]. Here we simply note that a crucial
element of maps quantized in this manner is that during one time step
vertical rectangles go into horizontal ones. Quantum mechanically this
corresponds to the partitioning of the state space into disjoint sets
in either the position or in the momentum basis. In reduced units
(when the position and momentum can range from 0 to 1), Planck's
constant $2\, \pi \, \hbar$ is $1/N$.  The classical limit is reached
as $N \rarrow \infty$. We observe that in this classical limit,
contrary to the usual one, the phase space remains finite. The
operator $B_{1}$ has the classical map's reflection symmetry about the
center of the square, the $R$-symmetry.  The operator $B_{1}$ commutes
with $R_{N}$, where $R_{N}$ is defined as

\beq \br n|R_{N}|n^{ \prime} \kt = \delta(n+ n^{ \prime}+1) mod(N) .\eeq

Here $|n \kt $ and $| n^{ \prime} \kt $ are position eigenvectors and
$n,\,n^{ \prime}\,=0,1,...N-1 $.  This results in the eigenfunctions
having an odd or even parity. In accordance with the rule of
performing statistical analysis on only one symmetry class, we will
separate out the eigenvalues corresponding to each symmetry class,
using the method suggested by Saraceno in [8].  The $T$ symmetry of
the classical map is an anticanonical time reversal symmetry resulting
from an interchange in position and momentum with a step backward in
time ($p\leftrightarrow q, t\rarrow -t$). The quantum map inherits
this symmetry as an antiunitary symmetry
\beq G_{N}B_{1}G_{N}^{\dagger}=(B_{1}^{-1})^{\star}.\eeq
All this is in complete parallel to the usual baker's map
quantization. In fact all the symmetric maps we consider are
such that their quantum propagators commute with $R$-symmetry,
and all the maps we consider (including the unsymmetric ones)
have the antiunitary $T$-symmetry.

The quantization of the other maps is also
straightforward. For example $S_{2}(1/5)$ has a quantum mechanical
propagator given by

\beq B_{2} = G_{N}^{-1} \left( \begin{array}{ccccc}
        	        G_{N/5} & 0 & 0 & 0 & 0\\
		        0 & G_{N/5} & 0 & 0& 0\\
			0 & 0 & I_{N/5} & 0 & 0\\
			0 & 0 & 0 &G_{N/5} & 0\\
			0 & 0 & 0 & 0 & G_{N/5}
	 \end{array}
           \right). \eeq

 The above propagator has the correct classical symmetries and classical limit.

\subsection{Eigenangle distributions}
\hs{.2in} Since the propagators associated with maps are unitary operators the
eigenvalues are specified by eigenangles. The eigenangles of
$S_{1}(1/3)$ are mostly irrational multiples of $2\, \pi$. The
rational multiples correspond to eigenangles of $0, \pi/2, \pi$ and $
3 \pi/2$  and these states are degenerate.  We are interested in the
sequence of maps $S_{1}(a)$ with increasing $a$, when such
degeneracies tend to vanish, and most of the states are of a mixed
nature.

  We recall that the spectral properties of quantum dynamical systems
have centered around the nearest neighbour spacing distribution which
is usually Wigner-like if the corresponding classical system is
completely chaotic and Poisson-like if it is integrable [9,16].  The generic
case of mixed dynamics with a measure of chaos and regularity
coexisting has not received such a clear answer.  While we do not
expect $S_{k}(1/l)$ or $S_{1}(a)$ to provide us with  such a case, as their
dynamics is not an {\em interlacing} of elliptic and hyperbolic
periodic orbits, it does have a set of measure zero chaotic orbits
whose effect at a {\em small but finite Planck's constant is to mimic
a transition from the Poisson like distribution to an intermediate one
as $k$ or $a$ is increased.}

In fig.3(a)(b)(c)(d) we show the nearest neighbour spacing (nns)
distribution for $S_{k}(1/2k+1), k=1,2,3,4 $ respectively.  The
slightly different values of Planck's constant arises since $N/2k+1$
must be an integer.  The transition from a Poisson-like (level
attraction) distribution when $k=1$ to an intermediate statistics
(level repulsion) when $k=4$ is apparent. {\em This suggests that the
underlying hyperbolic measure zero set of chaotic orbits is
contributing to the quantal spectrum}. The density of states can be
obtained as the time Fourier transform of the trace of the powers of
the propagator [9]. {\em The autocorrelation function therefore must show
the Cantor set becoming more prominent as the statistics tends to the
Wigner-like one}. Below we present some results on this.

 The fractal set of chaotic orbits affects the quantal spectrum,
reflecting  the fact that the behaviour of orbits  of classical short
periodic orbits strongly influences properties  of quantum mechanical
stationary wavefunctions. The chaotic set of measure zero in
the classical model takes an infinite time to develop.  The small but
finite value of the Planck's constant makes the {\em relevant time
scales finite}. The nns for the $S_{1}(a)$ map is shown in fig(4a),
for $N=300, a=149/300$. The reflection symmetry of the map permits the
statistical analysis of only 150 of them. The level repulsion in the
spectrum is apparent.

We may estimate {\em relevant time scales} by noting that there exist
regions of the phase space that are thin strips along the position or
momentum, of width $3^{-n}$ such that till $n$ time steps they undergo
only expansion and contraction but no rotation. Thus we may esimate
that at least until $n=Log_{3}(N)$ a {\em ``fattened''} set that
contains the {\em fractal set} must function as a hyperbolic set for
the purposes of quantal calculations (a ``pseudo-hyperbolic region'').
By the same arguments the parabolic periodic islands of period
$4(m+1)$ are not all resolved in quantum mechanics.  Recall that the
$\gamma_{m}$ island has an area of $3^{-m-2}$.  Periodic islands of
period $4(M+1)$, where
\beq  M\, >\, Log_{3}(N)\,-\,2 \eeq
cannot therefore be expected to function as such in the quantum
mechanics.  They cannot for instance be expected to appear in
autocorrelation functions for high times and cannot probably support
quantum eigenstates. In our discussion of the eigenfunctions we
will note that this naive argument gives us only a rough
estimate at best.

We noted in the section on classical mechanics that the breaking of
the $R$ symmetry produces in most cases a seeming transition to
ergodicity and chaos on finite measures. The map $SR$ in particular
was completely chaotic. We show the nns of the corresponding quantum
models in fig.4(b)(c)(d). Fig(4b) is the nns for the map $SRS$ with
partitions at $(0.3,0.6)$, with $N=300$;  fig(4c) is for partitions
at $(.3,.5)$ and $N=300$. Fig(4d) is the nns for $SR$ for N=300. In
this last map however we allowed a thin third strip  of width
1/300, making it strictly speaking a $SRS$ map. This was
done to preserve the toroidal boundary conditions.
The propagator chosen was thus
\beq B_{SR}=G_{N}^{-1}\left( \begin{array}{ccc}
			  G_{N/2}& 0&0\\
                          0&I_{N/2-1}&0	\\
			0&0&i
                         \end{array}
		\right), \eeq where $i$ is the square
root of $-1$.

 The autocorrelations not shown here indicate the ``closeness'' of these
maps, which may find a more quantitative meaning in the concept of
structural stability. By closeness we  imply visual closeness
of autocorrelation functions. {\em It is not clear how toplogical
conjugacy carries over into quantum maps}. Note that we have already
quantized a set of maps topologically conjugate to each other, and
found different statistics for the eigenangles. What is needed is a
sensible specification of distance in the space of quantum maps.

\subsection{Autocorrelation functions}

\hs{.2in}  We use the coherent state representation for finite dimensional
vector spaces, as developed by Saraceno in
[8].  The displayed function is a contour plot of $|\br p q|B^{n}|p q
\kt |$, with the $B$ being the appropriate quantum propagator.
Usually the square of this function is plotted as it is the
probability that a wavepacket initially localised at $(q,p)$, in the
sense of a minimum uncertainty packet, returns to its launching point;
The square being the quantal auto-correlation function at the point
$(q,p)$ at time $n$. We have however plotted just the absolute value
to emphasize fine structures. Fig.(5a) shows the autocorrelation for
$n=2$ and $N=48$ and for the operator $B_{1}$ with $a=1/3$, i.e the
quantum $S_{1}(1/3)$.  The maxima should occur at the classical
periodic points of period $2$.  The center of the square is a elliptic
fixed point and is so highlighted also as a period two orbit. The
corners of the square $(0,0)$ and $(1,1)$ are also fixed points
strictly belonging to the fractal hyperbolic set. The other two maxima
are due to the period two orbit $(1/4,\,3/4),\,(3/4,\,1/4)$
corresponding to the bi-infinite sequence $...020202.020202...$ ( or
$\ul{02}.\ul{02}$ if we denote with an underline the infinite
repetitions of a string), and it's orbit, which is simply a pure left
shift.

{\em This particular periodic orbit belongs to the  fractal  set we have
described above, which is of measure zero}. The associated periodic
points act as if they were usual hyperbolic periodic points with
eigenvalues $(1/3,3)$, which implies a quantum contribution {\em
similar} to that of an usual hyperbolic periodic orbit. The infinite
returns of this orbit that would affect the quantum correlations would
however be {\em different} from the usual case.  Considering that these are
usually decreasing contributions, it is possible that {\em these Cantor set of
points can affect the quantal spectrum; and in fact they do, as we see
from the eigenangle distributions}.

The fig.(5b) shows the autocorrelation for the operator $B_{1}$, again
with $a=1/3$ and $N=48$ for $n=4$. As we expect we see maxima along
classical periodic orbits. The whole central region, the square
$1/3<q<2/3, \, 1/3<p<2/3$ is now filled with direct parabolic orbits
and contributes the most to the autocorrelation. The 12 primitive
period 4 points belonging to the hyperbolic set are much smaller peaks,
but they are still highlighted in the figure. They correspond to the
orbits of the sequences $\ul{0220}.\ul{0220}, \, \ul{0002}.\ul{0002}$
and $\ul{2220}.\ul{2220}$. For example the peak at (14.4, 14.4)
corresponds to the periodic point $\ul{0220}.\ul{0220}$ whose
numerical value in base 10 is (3/10,3/10) which after scaling (multiplying by
48, the value of $N$) gives us (14.4,14.4). The primitive orbit of
period 2 noted above is also visible in this figure, with of course a
much reduced recurrence. The peaks around the fixed points are also
seen. The period 4 point (1/2,1/6) along with it's orbit contributes
significantly to the autocorrelation and is seen after scaling at the
point (24,8). These maxima around the central square are due to
inverse parabolic orbits that sit at the center of the period 8
periodic islands. Thus we see that much of the structure of the figure
can be explained in terms of the classical periodic orbits.

 We notice, however, the peaks at the center of the four edges that have
no periodic orbit support. They correspond on the classical torus to the
two points (0,1/2) and (1/2,0), which belong to the hyperbolic fractal
set.  These points are {\em homoclinic } to the fixed point at the origin and
are not near any period 4 orbit. For visualising the possibility of
such a recurrence it is helpful to think of the fate of a localised
distribution on the classical map centered at say (0,1/2). Under the
action of the map the distribution tends to elongate along the $q$
axis and contract along the $p$ axis, and as a whole move closer to
the $q$ axis and away from the $p$. At these time scales the fixed
point at the origin acts as an ordinary hyperbolic point. After 3 time
steps the distribution is sufficiently spread  to have a large enough density
at the bottom of the middle third region which after one more time
step gets {\em rotated} to the region (1,1/2); this on the torus being the
same as the point (0,1/2).

Such recurrences are known in general dynamical systems and have been
called ``homoclinic recurrences'' [10]. Their contributions to the
correlations are to be added to the periodic orbit recurrences and
presumably a complete semi-classical study of these maps would do so.
This would result in modifications to the periodic orbit sums for the
powers of the propagator, leading possibly to ``homoclinic sums''.  The
semi-classics of the usual baker's map has proven to conform to the
periodic orbit sum theory [9] {\em without} any homoclinic recurrences [10],
[11].  The value of the $SRS$ maps in regard to these comments depends
upon whether a nontrivial semi-classics exists. Our results here
indicate at least the presence of an interesting finite Planck's
constant domain and the discussions here suggest that this may be
possible for maps with a smaller rotating region, or with more
partitions. {\em The map $SR$ is completely chaotic and it remains to
be seen if the semi-classics of this map will also look like a
periodic orbit sum}.  The organization of the dynamics by the periodic
points of the usual baker's map may not be generic.

The classical topological conjugacy of the $S_{1}(a)$ maps has the
effect that the quantum autocorrelation functions have similar
structures, as is apparent on comparing figs.(5a) and (5c). The maxima
of the autocorrelations about the hyperbolic Cantor set of fixed
points is however increasing because the classical instability of
these orbits is {\em decreasing}. The higher hyperbolic contribution
coupled with a much lower contribution from the dominant parabolic and
elliptic regions contributes to the spectral transition.  Fig.(5d)
shows the autocorrelation $|\br pq|B_{2}^{2}|pq \kt |$, for $N=70$.
The orbits highlighted as maxima are
\begin{enumerate}
\item the five fixed points
$(\ul{0}.\ul{0},\,\ul{1}.\ul{1},\,\ul{2}.\ul{2},\,
\ul{3}.\ul{3},\, \ul{4}.\ul{4})$ and
\item the 12 period two orbits composed of the
$6$ possible combinations of the symbols $0,1,3,4$,
\end{enumerate} all of which
belong to the chaotic fractal set, except the center. The number of
such contributions increase with the number of partitions. This
generates increasing hyperbolic contributions to the autocorrelations,
 inducing in turn the spectral change shown in fig.(3).{\em  A
semi-classical theory of such maps should thereby shed light on the so
called mixed systems where chaos and order coexist}.

\subsection{Eigenfunctions}

\hs{.2in} Eigenfunctions of quantum systems with mixed or
chaotic classical limits are still in the process of being understood.
The phenomenon of periodic orbit scarring [12] has been observed in
several quantum eigenfunctions when the corresponding classical system
is chaotic. A semiclassical theory towards the understanding of this
phenomenon has been developed in [13] and [14]. Once again our maps
provide a complete collection of eigenfunctions for a mixed system
that is completely understood. Fig.(6) are the intensities of some
eigenfunctions of $B_{1}$, which is the quantum $S_{1}(1/3)$ map, for $N=48$.
The plot is a contour of the function
\beq
       |\br p\,q| \psi \kt |^{2}
\eeq
where $\br p\,q|\psi \kt$ is an eigenfunction of $B_{1}$ in the
coherent state representation. Fig(7) shows some classical periodic
points of this map belonging to the chaotic measure zero set that scar
some of the eigenfunctions shown in fig(6). The ternary code of these
orbits have also been indicated.

 Of the $48$ eigenfunctions $9$ representatives have been selected to
show classical structures. Of the remaining, many more are similar and
many do not apparently possess any prominent classical structures. A
majority of the eigenfunctions are concentrated around either the
regular parabolic and elliptic periodic orbits or are scarred by
periodic orbits belonging to the hyperbolic fractal set. We use
scarring  in the conventional way to indicate the presence of
classically ``improbable orbits''.  Figs.6(a)(b)(c)(d) are some
eigenfunctions {\em scarred by periodic orbits of the hyperbolic fractal
set}. Fig.(6a) shows an eigenfunction scarred by the period $2$ orbit
corresponding to the sequence $\ul{02}.\ul{02}$. Fig.(6b) shows the
scarring from two period $3$ orbits represented by the sequences
$\ul{202}.\ul{202}$, and $\ul{020}.\ul{020}$. Fig.(6c) shows high
intensities at the period $4$ orbit $\ul{2002}.\ul{2002}$.  Fig(d)
shows an eigenstate scarred by an orbit of as high a period as $6$,
corresponding to the sequence $\ul{220002}.\ul{220002}.$

As is expected, a large number of the eigenfunctions are
dominated by the parabolic islands and elliptic orbits. The trivial
central region of the classical phase space forms the support of four
eigenstates and fig.(6e) shows one such state. Fig.(6f) shows a state
dominated by the period $8$ parabolic island at $(1/3<q<2/3, \;
1/9<p<2/9)$ and it's $3$ partners. Fig.(6g) shows two of the three
island chains of period $12$. They correspond to the sequences $A
120.1B$ and $A 102.1B$ and their island partners. Here $A$ and $B$ are
arbitrary strings, indicating that the sequence represents an area
rather than a single point. Each island chain of this type has three
islands and are all highlighted here. The missing period $12$ island
chain from this eigenstate is the one consisting of $6$ islands with
one of the representatives being $A100.1B$.

In fig.(6h) we show an eigenstate which has prominent structures
apparently not related to any classical periodic orbit. In fact such
peaks along the edges of the square were observed by Saraceno [8] for
the quantum baker map. The prominence of hyperbolic trajectories
surrounding a hyperbolic fixed point has been observed in the
eigenfunctions of other systems, like the standard map. {\em The presence
of such scars in the eigenstates is once more an indication that for
the purposes of quantum mechanics the classical map seems to possess a
``pseudo'' or effective hyperbolic region of non-zero measure that we
have discussed above}. The prominent peaks at the centers of the edges
are probably due to the presence of a homoclinic recurrence that was
noted in the autocorrelattions.  Fig.(6i) shows an eigenstate that seems
to have no apparent classical support, although it may well correspond
to an interference of several longer periodic orbits. This state represents
a mixed state, while the majority of the eigenstates are scarred either
by the hyperbolic set or dominated by regular structures, but not both.
This shows that in such a system the seperation of states into regular
and chaotic subsets may be possible.

Thus the simple map $S_{1}(1/3)$ has a rich eigenstructure, with a
mixture of states dominated by the hyperbolic fractal set and by
states dominated by regular parabolic islands, and elliptic orbits. As
we decrease the area of rotation, either with the $S_{1}(a),\;a>1/3$
maps or the $S_{k}(1/2k+1)$ maps we would expect the proportion of
eigenstates with support on the hyperbolic set to increase. These
features may enable us to study the changes in the eigenstructures of
such systems tending towards global chaos. We finally note that the
periodic orbits of the fractal set that have scarred the
eigenfunctions have periods as high as $6$. At this stage in the
development of the fractal set the hyperbolic strips are of width
$3^{-6}$ or 1/729 which is very small compared with the value of
Planck's constant used, namely 1/48.  This shows that the quantum map
is able to resolve structures that are classically fine.  The exact
nature of the ``smoothing'' of the phase space due to the finite value
of Planck's contant is not fully understood. Recent work by P.W. O'
Connor et. al [10] indicates that in the quantum baker's map accuracy
of the semiclassical traces was observed to be good much beyond the
Log time [15], implying that the quantum dynamics explored delicate
structures which appeared, however, {\em not in the classical phase-space,
but in a mock phase-space} used to display objects in the coherent
state representation. Similar mechanisms may be causing here the
resolution of such delicate structures.

\section{Conclusions}

\hs{.2in} In this note we have introduced and studied some maps
of the square onto itself. The purpose has been to quantize the
simplest possible maps in which regular and irregular orbits coexist.
We have found that even a measure zero set of chaotic classical orbits
influences the quantal spectra and the eigenfunctions, due to the
finiteness of Planck's constant. Thus we observe a spectral
transition as when a system's phase space acquires a {\em measurably}
increasing chaotic component. In some of the maps considered here, the
class called $S_{1}(a)$, such a transition occurs even though the
underlying chaotic fractal set is of Lebesgue {\em measure zero} for
all $a$.  This interesting behaviour arises because we are looking at the
stationary states at a given Planck's constant.

The formulae for intermediate statistics as developed by Berry and
Robnik [17] depend on the ratio of measures of the chaotic and regular
regions of the phase space. This work was for the eigenvalues of a
continous time Hamiltonian system. For the class of maps $S_{1}(a)$
that exhibits a spectral transition such a ratio is of course zero.
However the arguments developed above suggests the existence of a
``pseudo-hyperbolic'' region of positive measure that depends on the
Planck's constant.  Part of an ongoing work is to verify whether such
an area which is culled from partly classical and partly quantum
conditions will produce statistics that agrees with the Berry-Robnik
formulae.  We have observed the scars of periodic orbits belonging to
this fractal set on the eigenfunctions which we have shown in a
coherent state  representation.

There are several other maps that form the general family we have
called the $SRS$, and their classical behaviour has proven to be
interesting. It remains to be seen if such a transformation which
introduces a discontinous rotation component to the well known baker's
map has any significance in the context of general dynamical sysytems.
The extreme case we have called $SR$ in which only one half of the
phase space is stretched, while the other is rotated has been shown to
be chaotic. It's quantum features follow those of completely chaotic
systems such as those of the usual baker's map.

\begin{center}
{\bf Acknowledgments}
\end {center}
\hs{.2in} The term ``Lazy Baker'' was first used in the context of
$SRS$ maps in the thesis of Dr. Ali Amiry. We are  grateful to him
for the first study of these maps and for several discussions.
A.L wishes to thank the Department of Physics, State University of New York
at Stony Brook and the National Science Foundation for their support.
This work was in part supported by the National Science Foundation.

\newpage
\appendix

\begin{center}
  {\bf Appendix A}
\vspace{.4in}
{\bf $SRS$ Map}.
\end{center}

\hs{.2in} Here we prove some statements of the text. In this
section of the appendix arbitrary infinite strings of 0,1 and 2 are
denoted by $A$ and $B$.  Finite strings of length $m$ comprising of 0
and 2 we denote by $\gamma_{m}$.  The operations of 2's complement and
``conjugation'' were defined in sec. 2.1. They are denoted by an
overbar and dagger respectively. An underline denotes infinite
repetitions of that string. All points in the square are assumed to be
in base 3.  First we consider the map $S_{1}(1/3)$.

Let $T$ be the unit square and the set $S$ be
\beq  T \setminus (C_{1/3} \times [0,1] \bigcup [0,1] \times C_{1/3}). \eeq
we first prove that all points in $S$ are periodic.
If  $x_{0}\in S$ then  its orbit must at some time $n$ be given by

\beq x_{n}=A1.a_{1}a_{2}\ldots a_{m}1B \equiv A1.\gamma_{m}1B
;\,\, a_{i}\in \{0,2 \},\,\, i=1,\ldots, m.\eeq

Here $m=0,1,2\ldots$. Then future orbit of $x_{n}$ is as follows:

\beq x_{n}=A1.\gamma_{m}1B \longra A1\gamma_{m}.1B \rarrow B1.\bar{\gamma}_{m}^
{\dagger}1\bar{A}
\longra B1\bar{\gamma}_{m}^{\dagger}.1\bar{A}\rarrow
\bar{A}1.\gamma_{m}1\bar{B} \eeq

\[ \longra
\bar{A}1\gamma_{m}.1\bar{B} \rarrow \bar{B}1.\bar{\gamma}_{m}^
{\dagger}1A
\longra \bar{B}1\bar{\gamma}_{m}^{\dagger}.1A \rarrow
vA1.\gamma_{m}1B =x_{n} \]

We have denoted by short arrows a single time step representing a
rotation, while a long arrow has $m-1$ {\em intermediate} states all
undergoing a simple left shift because the strings $\gamma_{m}$ and
$\bar{\gamma}_{m} ^{\dagger}$ have only 0 and 2 in them. This
completes the proof, as well as, it tells us that the period of $x_{n}$ is
$4(m+1)$, though there are isolated points of smaller primitive
periods as we will see below. Primitive periodic orbits of period
$4(m+1)$ are evidently continuos and are infinite in number. They are
of the direct parabolic kind. We can infer several other features from
the orbit of $x_{n}$,  written above. In particular, the point
$x_{n}$ will be of period (m+1) if $\bar{A}=B=A$ and
$\gamma_{m}=\bar{\gamma}_{m} ^{\dagger}$, where equality of two strings
means equality of each corresponding bits (We call such strings
$R$-symmetric strings). This implies that $A=B=\ul{1}$  and that $m$
must be even. These periodic orbits are rotated once and stretched $m$
times during a cycle, and hence are of the elliptic kind with the
following stability matrix:

\beq \left( \begin{array}{cc}
          0 & -3^{m}\\
     3^{-m} &0
     \end{array}
         \right ). \eeq

{\em The strangeness of the $SRS$ map is in this feature: stretching
along the $q$ direction is followed by a rotation which effectively
shifts the stretching to the $p$ direction which is then contracted!} We
note that the usual Baker compounded by a rotation of the whole square
by $ 90$ deg is not at all chaotic, it being period 4 as a whole! Of
course, such a process is aided by the symmetry of these maps, the
breaking of which seems to produce a whole class of mixed maps.

The elliptic orbits are therefore of only odd periods and if the
period is $(m+1)$ the number of such orbits is simply $2^{m/2}$ as
one half of the $\gamma$ string determines the other. Also from (23)
we see that if $\bar{A}=A, \bar{B}=B$ then the orbit is of period
$2(m+1)$ and this is a primitive orbit if $\gamma_{m}\ne \bar{\gamma}_{m}
^{\dagger}$. It follows that $A$ and $B$ are again simply strings
of 1. These orbits have repeated eigenvalues of $-1$ and are therefore
inverse parabolic;  given $m$ there are $2^{m}$ such orbits, including
the repetitions of period $(m+1)$ orbits.

We can also see the island structures from (23). Clearly the point
$A1.\gamma_{m}1B$ with no restrictions on $A$ and $B$  represent
a rectangular region defined by
\[3^{-m-1}<q<2.3^{-m-1},\,\, \, 1/3<p<2/3 ,\]
 having an area $3^{-m-2}$. The images of this rectangle will retain
the area and form a closed chain of period $4(m+1)$. The number of
such distinct island chains and the number of islands in each may also
be easily inferred. If $m$ is odd, none of the finite string are
$R$-symmetric. Since both the string and it's $R$-symmetric partner
appear in our specification of an island, there are $2^{m-1}$ distinct
island chains.  If $m$ is odd, there are two possibilites: a string may
or may not be $R$-symmetric. For a fixed $m$ there are $2^{m/2}$
strings of the former kind and therefore $(2^{m}-2^{m/2})$ of the
latter kind. We therefore get the total number of distinct island chains to be
\beq  2^{m/2} + (2^{m}-2^{m/2})/2 = 2^{m-1}(1+2^{m/2-1}).\eeq
The number of islands in each island is a simple consequence of the
above discussion.

\newpage

\appendix
\begin{center}
{\bf Appendix B}
\vspace{.4in}
{\bf $SR$ Map}.
\end{center}

\hs{.2in} We now prove that the map $SR$ is chaotic on the whole
measure. To get a complete picture of the position of the periodic
orbits and so forth we use the {\em binary} representation even though
the dynamics is a subshift of finite type on {\em three} symbols. In this
part of the Appendix $A$ and $B$ denote arbitrary strings of 0 and 1.
Any state in the square is represented as a bi-infinite string of 0
and 1.  Let the most significant bit of the position be 0, then the
dynamics is one of left shift. If the most significant bit is 1, then
it is an interchange of position and momentum combined with the
operation of exchanging 0 and 1 (1's complement) in the new position
(or old momentum). We note that, unlike the baker's map, the
doubly infinite sequence without the dot does not describe the
trajectory in the sense that it does not also serve as an itinerary.
For convenience we divide the square into the 4 equal open subsquares.
We denote by $R_{1}$ the left bottom square, by $R_{2}$ the left top
square, by $R_{3}$ the right top square and by $R_{4}$ the right
bottom square.  Any state in $R_{4}$ must traverse $R_{3}$ and $R_{2}$
in the following two time steps uniquely, these being merely
rotations.  This evident observation is an useful device when cast in
the following form:
\beq A0.1B \stackrel{\textstyle{2}}{\longra} \bar{A}10.\bar{B} .\eeq

The map's dynamics is one of ``eventual'' left shift with or without a
reflection about the center of the square. To see this we give several
rules that may be easily verified.  Represent an arbitrary initial
state as
\beq A.\gamma_{m}B ;\;\; m=1,2,\ldots.\eeq
The $\gamma_{m}$ string is the first $m$ bits of the position, where
$m$ is arbitrary. Denote the number of the 0-1, 1-0 neighbours in the
string $\gamma_{m}$ by $n(\gamma_{m})$. Let the most significant bits
of position and momentum be $a_{1}$ and $b_{1}$. The rules are the following.

\beq A.\gamma_{m}B \stackrel{\textstyle{K}}{\longra} \bar{A}
\bar{\gamma}_{m}^{\dagger}.\bar{B},\eeq
if any one of the following conditions is true:
 \begin{enumerate}
\item $a_{1}=0$ and $n(\gamma_{m})$ is odd. Then $K=2n(\gamma_{m})+m-1$.
\item $a_{1}=1$, $n(\gamma_{m})$ is even and $b_{1}=0$ . Then
$K=2n(\gamma_{m})+m+1$.
\end {enumerate}

\[   A.\gamma_{m}B \stackrel{\textstyle{K}}{\longra} A \gamma_{m}.B,\]
if any one of the following conditions is true.

\begin{enumerate}
\item $a_{1}=0$ and $n(\gamma_{m})$ is even. Then $K=2n(\gamma_{m})+m-1$.
\item $a_{1}=1$, $n(\gamma_{m})$ is odd and $b_{1}=0$ . Then $K=2n(\gamma_{m})
+m+1$.
\end {enumerate}
Here $K$ is the number of {\em intermediate} steps.  We do not need to
specify the rules for the case when the initial point is at $R_{3}$ as
it's forward iterate falls in $R_{2}$ and it's backward iterate in
$R_{4}$, both of which are covered above. Thus we see from the rules
that $SR$ leads to a destruction of information with time, one of the
key ingredients of deterministic chaos. The time needed to shift $m$
bits in the usual baker map is $m$;  here we see this time increasing
depending upon the number of 1-0, 0-1 neighbours that have been
shifted. This is a measure of the map's ``laziness''.

One may now prove the existence of a dense orbit. The contruction is
essentially identical to that of the usual baker's map. Consider the
point $x_{0}$ in $R_{4}$ such that its position bits are strings of
all possible permutations of 0 and 1 of all possible lengths. This is
usually constructed by first placing the permutations of a string of
length 1, then 2 and so on. Given any point $y$ in $R_{1}$ or $R_{4}$
one may approximate it to any arbitrary degree by finite length
strings. Say $y=\alpha.\beta$ is the arbitrary point and its
approximation is $\alpha_{L}.\beta_{L}$ consisting of the first L
significant bits.  Then the point $x_{0}$ has somewhere in its
position string the substring $\alpha_{L}\beta_{L}$. Since the last
bit in the string $\alpha_{L}$ of length L ({\em i.e.} the most
significant bit in the momentum of $y$) is 0 the number of 1-0, 0-1
neighbours is necessarily odd.  Therefore, from the above rules the
orbit of $x_{0}$ must at a certain time be at
$\ldots\alpha_{L}.\beta_{L}\ldots$. Thus the orbit of $x_{0}$ visits
any small neighborhood of any point. This orbit is thus dense in
$R_{1}$ and $R_{4}$. Therefore it is dense on the whole square since
the regions $R_{2}$ and $R_{3}$ are accessed from $R_{4}$ by a mere
rotation. The existence of a dense orbit proves the ergodicity of the
map $SR$.

We also briefly discuss periodic orbits of this map. Evidently there
are no periodic orbits lying wholly in the region $R_{1}$, thus it is
sufficient to count the periodic orbits in $R_{4} $. There is a
period 2 orbit, comprising of the points (1,1/2) and (1/2,1).  Apart
from these there are no periodic points on the boundary of the square.
Since the dynamics is one of eventual left shift, it must be rational.

In $R_{4}$ the orbit can be written as either
$\ul{\gamma_{m}}\,\,.\ul{\gamma_{m}}$ or
$\ul{\gamma_{m}}\,\,\ul{\bar{\gamma}_{m}}.\,\,
\ul{\gamma_{m}}\,\,\ul{\bar{\gamma}_{m}}$. In both cases $\gamma_{m}$ alone
determines the period. From the rules given above we find the time
period to be $T=2(n(\gamma_{m})+1) +m$. We allow $n(\gamma_{m})=
0,1,2,\ldots,m-1$. If $n(\gamma_{m})$ is even, the orbit is of the
second kind; if it is odd, it is of the first kind.

After some elementary combinatorics we
get the number of periodic orbits of period $T$, excluding the period 2 orbit
on the boundary, but including the fixed point, as

\beq p(T)=3(\sum_{j} \left( \begin{array}{c}
                      T-2j-3\\
			j

			\end{array}
			\right))+
		(\sum_{j} \left( \begin{array}{c}
                      T-2j-1\\
			j

			\end{array}
				\right)). \eeq
\[ \mbox {for} \;\;T>2,\;\; j=0,1,2,\ldots\]

The summations end when the combinatorial symbols become meaningless.
The above series for $p(T)$ satisfies the recursion relation
\beq p(T)\,=\,p(T-1)\,+\,p(T-3), \eeq  so that $p(T)$ follows a
Fibonacci like sequence

\[ 4,5,6,10,15,21,31,46,\ldots .\]
The first entry being the number of period three states and so on. The
stability matrix of a period $T=2n(\gamma_{m})+m+2 $ trajectory is

\beq        \left( \begin{array}{cc}
		2^{m+1}(-1)^{n} & 0  \\
		0 & 2^{-m-1}(-1)^{n}
			\end{array}
			\right).                                 \eeq
Hence all the periodic orbits are hyperbolic;  they are ordinary if
$n\equiv n(\gamma_{m})$ is even and they are reflecting if $n$ is odd.

We may note that (in keeping with the rather special features of the
above maps) the map $SR$ also shows some non-generic features although
it is chaotic on the whole measure. The stable and unstable manifolds
of a periodic point are known to foliate the phase space and to
usually intersect transversally. The stable manifolds of the fixed
point at the origin are the vertical lines in the regions
$R_{1},R_{2}$ and $R_{4}$ with the position being of the form
$k/2^{l}$, for any two integers $k$ and $l$. In the region $R_{3}$
they are {\em horizontal} lines along similar points. This is
understandable as the whole region $R_{4}$ is rotated into $R_{3}$,
but it leads to non transverse intersections of the stable and
unstable manifolds and two whole {\em intervals} of orbits homoclinic
to the fixed point, the intervals defined by the boundary between
$R_{3}$ and $R_{4}$ and the boundary between $R_{3}$ and $R_{2}$.

\newpage
\begin{center}
{\bf References}
\end{center}

\begin{enumerate}

\item {\sc J.H. Hannay and M.V. Berry}, {\sl Physica} {\bf 1D} (1980), 267.

\item {\sc F.M. Izrailev}, {\sl Phys. Rep.} {\bf 196} (1990), 301.

\item {\sc K. Falconer}, ``Fractal Geometry'', John Wiley \& Sons, England,
1990.

\item {\sc S. Grossmann and S. Thomae}, {\sl Z. Naturf. a} {\bf 32} (1977),
1353.

\item {\sc R.L. Devaney}, `` An Introduction to Chaotic Dynamical Systems,''
 Benjamin, Menlo Park, 1986.

\item {\sc M. Saraceno and A. Voros}, {\sl Preprint}

\item {\sc N.L. Balazs and A. Voros}, {\sl Ann. Phys. (N.Y.)} {\bf 190} (1989),
 1.

\item {\sc M. Saraceno}, {\sl Ann. Phys. (N.Y)} {\bf 199} (1990), 37.

\item {\sc M.C. Gutzwiller}, ``Chaos in Classical and Quantum Mechanics',''
Springer-Verlag, NewYork, N.Y, 1990.

\item {\sc P.W. O'Connor, S. Tomsovic and E.J. Heller} {\sl Physica}{\bf 55D}
(1992), 340.

\item {\sc A.M. Ozorio de Almeida and M. Saraceno}, {\sl Ann. Phys. (N.Y)}
{\bf 210} (1991), 1.

\item {\sc E.J. Heller} {\sl Phys. Rev. Lett.} {\bf 53} (1984), 515.

\item {\sc M.V. Berry}, {\sl Proc. Roy. Soc. Lond.} {\bf A423} (1989), 219.

\item {\sc E.B. Bogomolny}, {\sl Physica} {\bf 31D} (1988), 169.

\item {\sc M.V. Berry and N.L. Balazs}, {\sl J.Phys.} {\bf A12} (1979), 625.

\item{\sc O. Bohigas and M.J. Giannoni}, in ``Mathematical and computational
  Methods in Nuclear Physics'' (ed. by J.S. Dehesa, J.M.G. Gomez, and A. Polls
), Lecture Notes in Physics {\bf 209}, Springer Verlag, 1984.

\item{\sc M.V. Berry and M. Robnik}, {\sl J. Phys.} {\bf A19} (1986), 649.

\end{enumerate}

\newpage

\begin{center}
{\bf Figure Captions}
\end{center}

\begin{description}

\item[Figure (1)] Definition of the map $SRS$. Fig.(1a) and fig.(1b)
are pictures of the unit square  before and after the transformation.

\item[Figure (2)] Fig.(2a) shows 20,000 iterations of the point (0.2,0.2),
for cuts given by $a=.3$, $b=.5$. Fig(2b) shows 20,000 iterations of
the point (0.16,.5) for $a=.33,b=.66$, when the $R$-symmetry has been
just broken. This indicates the structural instability of $S_{1}(1/3)$.
Note the change of the periodic point at (1/2,1/6) and its orbit into
hyperbolic regions.

\item[Figure (3)] Nearest neighbour spacing distribution for the
quantum maps of $S_{1}(1/3), S_{2}(1/5), S_{3}(1/7)$, and $S_{4}(1/9)$
are shown in figs.(3a), (3b), (3c) and (3d) respectively; the
corresponding values of Plancks's constant, $(1/N)$ are 1/288, 1/290,
1/294, 1/288. The reason for choosing slightly different values of the
Planck's constant is explained in sec.3.2.

\item[Figure (4)] Fig.(4a) shows the nearest neighbour spacing distribution
for the symmetric map $S_{1}(a)$ with the rotating region being very
small; $a=149/300$, and $N=300$. Figs.(4b) and (4c) are the nns
distributions for the unsymmetric $SRS$ for $a=.3, b=.6, N=300$, and
for $a=.3, b=.5, N=300$ respectively. Fig(4d) shows the nns for the
map $SR$, for $N=300$.

\item[Figure (5)] Autocorrelation functions for the following quantum maps
for different times and $N$ values. $S_{1}(1/3)$ with $ n=2, N=48$ in
fig.(5a).  $S_{1}(1/3)$ with $n=4, N=48$ in fig.(5b). $S_{1}(11/24)$
with $n=2, N=48$ in fig.(5c).  Note that the similarity between this and
fig.(5a) is due to the classical topological conjugacy discussed in
the text.  $S_{2}(1/5)$ with $n=2, N=70$, in fig.(5d).

\item[Figure (6)] Eigenfunctions of the quantum $S_{1}(1/3)$
map, for N=48. Nine of the 48 are shown.

\item[Figure (7)] Some periodic orbits of $S_{1}(1/3)$ which belong
to the chaotic fractal set and which scar some of the eigenfunctions
shown in fig.(6).
\end {description}

\end{document}